\documentclass[pmlr,twocolumn,10pt]{jmlr} 

\usepackage{booktabs}
\usepackage{graphicx}

\usepackage{multirow}
\usepackage{color}
\usepackage{svg}
\usepackage{xcolor,colortbl}
\usepackage[disable]{todonotes}
\usepackage[T1]{fontenc}

\usepackage[group-separator={,}]{siunitx}

\definecolor{light-gray}{gray}{0.83}
\newcommand{\GREY}{\cellcolor{light-gray}\bf} 

\theorembodyfont{\upshape}
\theoremheaderfont{\scshape}
\theorempostheader{:}
\theoremsep{\newline}

\title[Window Stacking Meta-Models for Clinical EEG Classification]{Window Stacking Meta-Models for Clinical EEG Classification}

\author{%
\Name{Yixuan Zhu} \Email{yixuan2.zhu@live.uwe.ac.uk}\\
\addr University of the West of England, UK
\AND
\Name{Rohan Kandasamy} \Email{rohan.kandasamy@nhs.net}\\
\addr  University College London Hospitals NHS Foundation Trust, UK\\
\addr  UCL Queen's Square Institute of Neurology, UK
\AND
\Name{Luke J. W. Canham} \Email{luke.canham@nbt.nhs.uk}\\
\addr  North Bristol NHS Trust, UK
\AND
\Name{David Western} \Email{david.western@uwe.ac.uk}\\
\addr University of the West of England, UK
}
\begin{document}

\maketitle

\begin{abstract}

Windowing is a common technique in EEG machine learning classification and other time series tasks. However, a challenge arises when employing this technique: computational expense inhibits learning global relationships across an entire recording or set of recordings.
Furthermore, the labels inherited by windows from their parent recordings may not accurately reflect the content of that window in isolation.
To resolve these issues, we introduce a multi-stage model architecture, incorporating meta-learning principles tailored to time-windowed data aggregation. We further tested two distinct strategies to alleviate these issues: lengthening the window and utilizing overlapping to augment data. Our methods, when tested on the Temple University Hospital Abnormal EEG Corpus (TUAB), dramatically boosted the benchmark accuracy from 89.8 percent to 99.0 percent. This breakthrough performance surpasses prior performance projections for this dataset and paves the way for clinical applications of machine learning solutions to EEG interpretation challenges. On a broader and more varied dataset from the Temple University Hospital EEG Corpus (TUEG), we attained an accuracy of 86.7\%, nearing the assumed performance ceiling set by variable inter-rater agreement on such datasets.
\end{abstract}

\paragraph*{Data and Code Availability}
Our study includes electroencephalography (EEG) datasets collected from \url{https://isip.piconepress.com/projects/tuh\_eeg/}. Our code is shared on \url{https://github.com/zhuyixuan1997/EEGScopeAndArbitration}.

\section{Introduction}
\label{sec:intro}

\subsection{Background}
\label{sec:background}

Electroencephalography (EEG) plays a pivotal role in diagnosing, monitoring and prognostication in a plethora of neurological conditions, but there is a paucity of trained EEG reporters globally. A cornerstone of clinical EEG analysis lies in discerning between normal and abnormal recordings. Over the years, the application of machine learning for this purpose has gained traction, with many studies delving into its potentials \citep{schirrmeister2017deep,amin2019cognitive,banville2021uncovering,banville2022robust,gemein2020machine,muhammad2020eeg,wagh2020eeg,alhussein2019eeg,roy2019chrononet}.

Prominent research in this domain predominantly relies on the Temple University Hospital Abnormal EEG Corpus (TUAB) for both training and assessment \citep{lopez2017automated}, which is an annotated subset of the larger Temple University Hospital EEG Corpus (TUEG) \citep{obeid2016temple}.

Since the inception of the Deep4 convolutional neural network in 2017 \citep{schirrmeister2017deep}, advancements in machine learning accuracy for this task, when gauged on TUAB, have been incremental at best --- moving from 85.4 percent with Deep4 to a high of 89.8 percent \citep{muhammad2020eeg}, as detailed in \tableref{tab:models}. \citet{gemein2020machine} even conjectured the presence of an accuracy ceiling close to 90 percent, drawing parallels with established inter-rater agreement levels among human experts in routine clinical settings.

In our study, we argue that previous achievements were not confined by the original label precision, but rather by their application during data preparation for machine learning. One often overlooked distinction between traditional clinical practice and the majority of deep learning methodologies is the application of the `normal/abnormal' designation to an entire EEG session, corresponding to a single clinical appointment.
In a standard clinical scenario, professionals determine abnormal brain activities by analyzing the entirety of the data in the session, leading to one comprehensive label for the session, or at least consider non-specific abnormalities in the wider context of the record.
However, in many contemporary machine learning paradigms, direct integration of a complete recording into a model is computationally challenging. This is largely due to the fact that a vast input vector would require an equally expansive parameter set in the model.
To circumvent this, recordings are fragmented into more manageable windows, enhancing the overall training sample size.
During the training phase, each of these windows assumes the label of its originating recording. Evaluations are typically conducted at the recording level by pooling the classifier outputs from individual windows.
This process, which we term `arbitration', is introduced for clarity and to distinguish it from other aggregation methods, such as the temporal consolidation of a model's input attributes as studied by \citet{gemein2020machine}.

\citet{western2021automatic} observed that such inheritance of labels from broader sessions could be misleading. A session designated as `abnormal' might contain isolated abnormal graphoelements, yet several windows within it might be devoid of abnormalities. When used for training, such windows are still tagged `abnormal'. Consequently, a model might identify them as `normal'. The ultimate verdict, based on aggregation, could then be skewed, resulting in diminished sensitivity and a rise in false negatives. This hypothesis finds support in \tableref{tab:models}, which indicates a consistent trend of subdued sensitivity across leading models, hinting at the prevalence of this issue.

Following this observation, in a preliminary report \citep{zhu2023scope} we validated the above hypothesis by demonstrating that noteworthy improvements in accuracy --- and more specifically sensitivity, a notable weakness of prior studies outlined in \tableref{tab:models} --- could be obtained either by increasing window length or adopting a form of machine learning arbitration.
In the present study, we consolidate the latter concept in the more general form of window stacking meta-models.
Further contributions of the present study include:
\begin{itemize}
\item
optimisation of first- and second-stage architecture choices to further improve upon state-of-the-art accuracy on TUAB.
\item
evaluation of the benefits of overlapping windows in the context of a conventional approach as well as window stacking meta-models.
\item
extension of the concept from recording-level to session-level arbitration with a third-stage model.
\item
evaluation against a larger and more diverse dataset, with labels extracted from clinical reports using natural-language processing. Here the method of \citet{western2021automatic} is improved with the use of a weighted loss function, avoiding the need to balance the dataset by omitting samples.
\item
examination of algorithmic explainability of the proposed meta-models.
\end{itemize}

\begin{table}[htbp]
\floatconts
    {tab:models}
    {\caption{Summary of state-of-the-art performance metrics for different models applied to abnormal EEG classification on the TUAB dataset
    }}%

    \resizebox{\linewidth}{!}
    {
    \begin{tabular}{|c|c|c|c|c|}
    \hline
    \GREY Model  &  \GREY Accuracy & \GREY Sensitivity & \GREY Specificity \\\hline
    1D-CNN (T5-O1 channel) \citep{yildirim2020deep} & 79.3 \% &71.4  \% &  86.0 \%   \\\hline
    1D-CNN (F4-C4 channel) \citep{yildirim2020deep} & 74.4 \% &55.6  \% & 90.7 \%   \\\hline
    Deep4 \citep{schirrmeister2017deep}     & 85.4 \% & 75.1 \%  & 94.1 \%  \\\hline
    TCN \citep{gemein2020machine} & 86.2 \% & &    \\\hline
    ChronoNet \citep{roy2019chrononet} & 86.6 \% & &    \\\hline
    Alexnet \citep{amin2019cognitive}  & 87.3 \% & 78.6 \%  & 94.7 \%  \\\hline
    VGG-16 \citep{amin2019cognitive}  & 86.6 \% & 77.8 \%  & 94.0 \%  \\\hline
    Fusion Alexnet \citep{alhussein2019eeg}	& 89.1 \%  & 80.2 \%&  96.7 \%  \\\hline
    Fusion CNN \citep{muhammad2020eeg}	& 89.8 \% &81.3 \% & 96.9 \%   \\\hline
    Scope and Arbitration (Deep4-ANN-Hybrid) \cite{zhu2023scope} & 93.3 \%   &	92.0 \%   &	92.9 \%   \\\hline
    \bfseries{Window Stacking Meta-Model (TCN-XGBoost-Raw)} & \bfseries{99.0} \%   &	\bfseries{98.1} \%   &	\bfseries{100} \%   \\\hline

    \end{tabular} 
}
 
\end{table}

\subsection{Proposal}
\label{sec: pro_arb}

As noted in our preliminary report \citep{zhu2023scope}, virtually all recent state-of-the-art machine learning approaches to this task apply some windowing to the original recording.
Some form of arbitration is then used to combine per-window outputs into a single decision, as depicted in \figureref{fig:EEG_classification}.
\citet{alhussein2019eeg}, and later \citet{muhammad2020eeg}, avoid this by fusing higher-dimensional features across multiple windows to produce a per-recording classification from an end-to-end trained (i.e. single-stage) architecture.
However, this approach significantly increases the number of model parameters and simultaneously reduces the number of training samples available (compared with per-window training), exacerbating the data-scarcity challenge in this task.

\begin{figure}[htbp]
\floatconts
  {fig:EEG_classification}
  {\caption{Generic diagram of a typical deep learning approach to clinical EEG classification.
  }}
  {\includegraphics[width=1\linewidth]{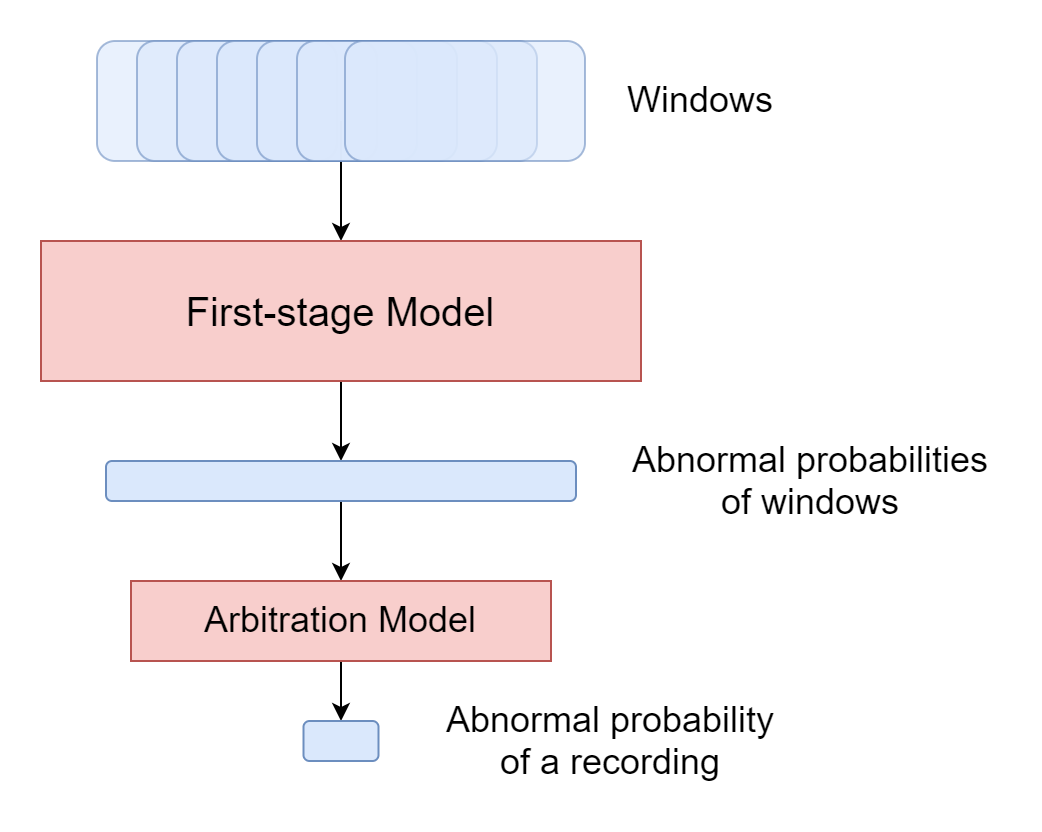}}
\end{figure}

Otherwise, the pervasive arbitration stage is not substantially discussed in prior literature.

In their examination of the impact of recording duration on outcomes, \citet{schirrmeister2017deep} consistently employed 60-second windows. 
While the arbitration methodology was not stated in that publication, an analysis of their source code reveals that the mean was taken across the outputs of each window. 
Subsequently, the parent recording was then categorized based on whichever mean probability was greater, either `normal' or `abnormal' (argmax). 
This use of `mean' differs from intuitive and conventional clinical practice, in which the presence of a temporally isolated abnormality (i.e. a single `abnormal' window) may justify classifying the recording as `abnormal' even if the majority of the windows are deemed `normal'.

By introducing machine learning at the arbitration stage, we sought to enable the overall architecture to optimise this decision, whether by adopting conventional human approaches, developing novel strategies, or any combination of these.
Furthermore, by training the arbitration stage separately we avoid the exacerbation of data scarcity problems that we perceived in the approach used by \citet{alhussein2019eeg} and \citet{muhammad2020eeg}, as well as reducing the computational expense of iterative development, thus enabling more thorough optimisation.

Although it has not been thoroughly considered for EEG classification, the notion of arbitration - or aggregation of per-window predictions - is not novel in machine learning.
Non-parametric approaches such as `mean' or `max pooling' are common, and fusion of features from multiple windows in a single end-to-end trained model, as used by \citet{alhussein2019eeg}, have also been explored in other domains. 
Furthermore, much attention is given to machine learning approaches for aggregation of predictions from ensembles of models, i.e. multiple models applied to the same data.
Such use of a second-stage machine learning model (meta-model) is typically referred to as `model stacking' \citep{wolpert_stacked_1992, pavlyshenko_using_2018}.
Our approach is similar except that it combines outputs from a single model applied to multiple windows.
Hence we refer to it as `window stacking'.

We are not aware of any prior works devoted to machine learning aggregation of predictions from multiple windows rather than multiple models. However, many more specific techniques used for aggregation among ensembles of models are likely to be applicable to arbitration between per-window predictions.
For example, \citet{susan_evaluating_2021} recently found XGBoost \citep{chen2016xgboost} to be effective as a meta-learner for an ensemble of models applied to an activity recognition task, while our results demonstrate its efficacy for arbitration between per-window predications.
We explore XGBoost as well as simple artificial neural networks (ANNs) as candidate architectures for the window stacking meta-model.

\section{Method}
\subsection{Data}

\subsubsection{Overview}

The TUEG dataset (Version: 2.0.0), compiled at Temple University Hospital (TUH) since 2002, offers a wealth of clinical EEG recordings based on the standard 10-20 electrode placement system. Comprising 14,987 patients, 26,846 sessions, and 69,652 recordings, TUEG is an unlabelled dataset; labels must be applied to enable its use for classifier training. 

\subsubsection{TUAB}

TUAB (Version: 2.0.0) \citep{lopez2017automated} is a subset of 2993 recordings from TUEG that have been labelled as normal/abnormal and divided into training and evaluation sets. The training set contains 1371 normal sessions and 1346 abnormal sessions. The test set contains 150 normal sessions and 126 abnormal sessions. Only one file from each session was included in this corpus. 
Similar to the approach in the Deep4 paper \citep{schirrmeister2017deep}, we removed the first minute from all recordings on the basis the signal quality is often poorer in this section. 
Because all examples in TUAB are over 15 minutes in duration, we did not impose any additional inclusion criteria based on the recording length. 
We used a maximum of 20 minutes from each recording.

\subsubsection{AutoTUAB}

While TUAB was used for experiments to refine our approach and for comparison with prior studies, to enable a more thorough evaluation we extracted a larger labelled dataset from TUEG.

This was achieved using automatic labelling based on natural language processing of text reports, as described by \citet{western2021automatic}.
We refer to this automatically labelled alternative to TUAB as `AutoTUAB'.
We selected only data with label confidence exceeding 99\%, length greater than 6 minutes, and all the 21 desired channels.
Following the screening process, AutoTUAB contained 26,504 recordings - 19,109 abnormal recordings and 7,395 normal recordings, across 18,747 sessions. 
It should be noted that, during the preparation of this manuscript, \citet{kiessner_extended_2023} introduced an improvement upon our auto-labelling approach, producing unbalanced (TUABEX) and smaller-but-balanced (TUABEXB) variants of an auto-labelled expansion of TUAB.

Although we did not have the opportunity to incorporate their approach, we present here a further improvement, which could enable wider use of their larger TUABEX dataset.
Like \citet{kiessner_extended_2023}, in \citep{western2021automatic} we omitted some records to achieve a balanced dataset.
In the present study, to accommodate the unbalanced nature of the data in AutoTUAB without reducing the dataset, we adopted a weighted loss function in the training of machine learning models.
Cross-entropy was used as the loss function, and the contribution of each sample was given a weighting $a_i$ that was inversely proportional to the number of samples $n_i$ in its class $i$.

\begin{equation}
a_i = \frac{\max(n_{normal}, n_{abnormal})}{n_i}
\end{equation}

Compared to TUAB, the AutoTUAB dataset is larger and more diverse. 
It is arguably more representative of clinical data since it is not manually selected, whereas the examples in TUAB were selected to form a dataset conducive to machine learning \citep{lopez2017automated}. Regarding the data partitioning of AutoTUAB, we adopted a $9:1$ ratio to divide the dataset into training and testing sets. To better evaluate the model's generalization ability, or in other words, to prevent information leakage from the test set to the training set, our division was based on patients rather than individual recordings. This ensured that all patients in the test set were not previously encountered in the training set, providing a robust test of the model's ability to generalize to new, unseen data.

\subsection{First-Stage Model}

In order to validate the efficacy of our proposed approach, we have selected four first-stage models, including two based on convolutional neural networks and two based on transformers. Specifically, the two convolutional neural networks are the well-established Deep4 model and the temporal convolutional network (TCN) \citep{bai2018empirical} as used by \citet{gemein2020machine}. Meanwhile, the two transformer models we have selected are the Vision Transformer (ViT) model \citep{dosovitskiy2020image} and the Patchout Audio Transformer (PaSST) \citep{koutini2021efficient}.

PaSST was originally designed for audio data and pre-trained on Audioset, a large set of 10-second audio clips extracted from YouTube videos \citep{gemmeke2017audio}, whereas we trained ViT solely on the EEG datasets described in this paper. 

With regards to pre-processing and hyperparameter settings, we have used the same pre-processing methods and settings as the original papers for Deep4 and TCN, while for ViT and PaSST we have used our own settings, as no prior precedent exists for their use on EEG data. Given that the data volume of TUAB and AutoTUAB makes it difficult to train large networks like ViT (even when using the smallest version), we have scaled down the model size. This includes reducing the dimensions of the key, query, and value, decreasing the dimensions of the MLP, and reducing the depth (i.e., using fewer multi-head self-attention modules). In addition, given that Vision Transformer (ViT) and PaSST were not originally designed for Electroencephalography (EEG) data, we have also modified the data preprocessing and initial layers of these two models to make them suitable for EEG classification tasks. 
ViT was applied directly to the time-series data, whereas for PaSST we used the Short-Time Fourier Transform (STFT) to convert raw data into time-frequency domain images.

Given that the performance of Temporal Convolutional Networks (TCN), PaSST, and Vision Transformer (Vit) significantly outperformed that of Deep4, all of our subsequent experiments concerning window length, window stride, and those performed on autoTUAB were conducted based on the premise of using Deep4 for the first-stage model. Full details of our implementation can be inspected in our open-source code repository.

\subsection{Second-Stage Models for Arbitration (Meta-Models)}
As stated in \sectionref{sec: pro_arb}, the purpose of the arbitration stage is to combine the per-window class probabilities into a single classification of the EEG session. 
Previous work does not discuss arbitration, although some form of arbitration is inevitable where models are applied to windows and evaluated on a per-recording basis (e.g. \citep{schirrmeister2017deep, gemein2020machine}). 
Inspection of the code used in the original Deep4 implementation \citep{schirrmeister2017deep} reveals that the `Mean' method was to integrate the results of windows. In some studies based on time-frequency images, transformations such as STFT are used to freely choose the size of the image \citep{alhussein2019eeg}, thus avoiding the need for windowing.

\subsubsection{Baseline}
In order to establish a comparison for the effectiveness of our proposed approach, we evaluated three baseline methods. 
\begin{description}
\item[No arbitration:] Here we simply evaluate the accuracy of predictions for individual windows compared with their label inherited from the parent recording. 
\item[Mean:] The `mean' method is that used in the Deep4 code \citep{schirrmeister2017deep} where the abnormal probability of all windows in a recording is averaged (arithmetic mean) and used as the abnormal probability of the entire recording. 
\item[Geomean:] This is the geometric mean of the abnormal probability of all windows in a recording. 
This choice is motivated by the fact the geometric mean is better suited to summarising multiplicative relationships, such as combined probabilities.
\end{description}

All three baseline methods are non-parametric, avoiding the risk of overfitting, and only integrate the window results based on some assumptions about the data. 
We used these baselines to evaluate whether a machine learning arbitration model can learn more complex mappings and improve performance.

\subsubsection{Pre-Processing of Meta-Model Inputs}

In our machine learning arbitration models, we utilized three distinct pre-processing approaches for inputs, derived from the first-stage model outputs: Raw, Histogram, and a Hybrid approach which combines elements of both. The Raw' approach processes individual recording results, padding with zeros for consistency. The Histogram' method employs a flexible representation using ten bins to handle varying recording lengths and window sources. The `Hybrid' method is a concatenation of the former two. For detailed explanations and visual representations, refer to our previous work \cite{zhu2023scope}.

\subsubsection{Meta-Model Architectures}

\paragraph{Overview}

We explore two meta-model architectures/algorithms in this study.
The first is an Artificial Neural Network (ANN), specifically a Multi-Layer Perceptron (MLP).
The second is XGBoost \citep{chen2016xgboost}.
To ensure the robustness of the results, five experiments were conducted for each first-stage model. For each first-stage instance, five experiments were carried out with each second-stage model architecture and hyperparameter setting . 
Thus, a total of 25 experiments were conducted for each architecture and hyperparameter setting when considering the two-stage architecture as a whole.

\paragraph{Artificial Neural Network (ANN)}

To optimize the performance of the ANN, a grid search was performed on the depth of the model (i.e., the number of hidden layers ranging from 0 to 3) and the number of nodes in the hidden layers (from 5 to 20).
Different activation functions (RELU, GELU, ELU, and None), Adaptive Pooling, and convolutional layers were also tested. 
However, they were not found to alter performance significantly \cite{zhu2023scope}. For simplicity, unless otherwise specified, we default to using a single-layer fully connected network followed by a softmax layer, without employing any other techniques.

\paragraph{XGBoost}

Recognising that the outputs of the first-stage model might be considered to be tabular data, we evaluated XGBoost \citep{chen2016xgboost} as an arbitration model, since it is frequently found to offer good performance on tabular data. 
Furthermore, it has recently proven to be effective as a meta-model for model stacking  \citet{susan_evaluating_2021}.
Within each training run, we conducted a grid search to select the optimum `maximum depth' parameter from the set $\{5, 10, 15, 20, 25\}$ based on three-fold cross validation.
All other hyperparameters were fixed at typical values based on prior experience.

\subsection{Third-Stage Models for Session-Level Arbitration}

Just as our meta-model at the second-stage aggregates classification results of windows within a recording, we consider the introduction of a third-stage model to aggregate all the results of all recordings within the session, using the output of the second-stage model as input and outputting the abnormal probability of the session. 
Since each session in TUAB's test set has only one recording, the experiment of the third-stage model is only carried out on AutoTUAB. 
However, since each session in AutoTUAB contains very few recordings, especially on the test set, with the number of recordings on the test set being only about 1.07 times the number of sessions, we restrict our search to the simpler non-parametric methods: `mean' and `geomean', which are consistent with the methods used in the second-stage model. 

In order to make a fair comparison between the results of the second-stage model (at the recording level) and the results of the third-stage model (at the session level), the results of the session are regarded as the results of all the recordings in it, thus converting the accuracy rate on the session level into the accuracy rate on the recording level.

Given that each session contains a relatively small number of recordings, we anticipated at the outset of the experiment that arbitration at the session level would not greatly enhance the overall performance of the model. Therefore, we took an additional experiments for sessions containing more than one recording. Then, we computed the confusion matrix again, subsequently calculating accuracy, sensitivity, and specificity. This allowed us to gauge the performance when the third-stage model is effectively in operation. This approach provides a more nuanced understanding of the model's performance when dealing with multiple recordings within a single session, which is closer to real-world application scenarios.

\subsection{Window Length and Overlapping}

We previously mentioned two issues in EEG classification tasks: one being the problem of misleading redistribution of labels, and the other being the limit imposed on model size by the amount of training data available. 
Our proposed meta-models effectively alleviate these issues. However, these challenges can also be addressed to some extent by simpler methods which can be combined with the multi-stage architecture to potentially achieve even better results.

The first approach involves increasing the window length, which allows each window to encompass a broader scope, thus improving the accuracy of the redistributed labels. 
We tested five different window lengths: 60, 180, 300, 400, and 600 seconds. The rationale behind this design is that, when using smaller windows, the remainder at the end of a recording is relatively small; thus the window lengths increase linearly from 60 to 300 seconds. In contrast, when using larger windows (from 300 to 600 seconds) with no overlap, we examine window lengths that are factors of the most common recording length (1200 seconds) to minimise the amount of information discarded. 
For each window length, we tested the performance of both the first-stage and second-stage models on the TUAB dataset.

On the other hand, we can employ overlapping windows (where the window stride is smaller than the window length) during windowing to preserve information and yield more data samples, thus mitigating the issue of data scarcity. 
We tested the effects of using overlapping windows in the case where window length is 60 s by applying strides of 10-60 s (83.3-0 percent overlap). In our experiment, overlapping was employed consistently in both the first-stage and second-stage models, meaning that both stages utilized the same window. This ensures that the data segmentation strategy remains consistent across different stages of our model, thereby maintaining the continuity of data representation and facilitating the comparison of performance between different stages of the model.
We then evaluated the performance of both the single-stage (`no arbitration') and two-stage architectures under these conditions.

\subsection{Exploring Information Distribution in Recordings}
\label{sec: info_dist}

When using machine learning for medical diagnostics or analysis, we often hope to gain human-understandable knowledge from the process, i.e. to learn from what the machine has learned. Although the decision-making process of deep learning models is difficult to interpret, the meta-models can provide insights into which windows are most important in it's decision making. 
When `raw' inputs are used for a single-hidden-layer ANN meta-model, the model's weights indicate the importance attributed by the meta-model to each window position $j$.
We calculate the importance $I_j$ by squaring the window weights (to account for the possibility of negative weighting) and averaging across both classes, as follows.

\begin{equation}
\label{eq:importance}
I_{j} = \sum_{i=0}^{1}\frac{W_{ij}^{2}}{2} 
\end{equation}

Here the meta-model's weights are denoted as $W_{ij}, i\in\{0,1\},0 \le j<T$, where $i$ is the category, $j$ is the window number (position in sequence), and $T$ is the maximum number of windows included in the recording.

\subsection{Using First-Stage Features in the Meta-Model}
As an extension of our proposal, rather than simply aggregating the outputs of the first-stage model, we consider the inclusion of features from intermediate layers of the first-stage model as additional inputs to the meta-model.
For simplicity, we restrict this experiment to using Deep4 as the first-stage architecture, with 60 s window length.

\figureref{fig:model diagram} presents a simplified diagram of this architecture. 
In the case of Deep4, the `Feature Extraction Layers' comprise the convolution blocks. We consider the `features' of length 26,800 extracted as the outputs of these layers as a candidate for a set of inputs to the meta-model. As another candidate, we consider the logits (2 per window) generated by the classifier layers. These are simply non-normalized variants of the probability estimates output by the SoftMax layer.

We then input these generated features into the second-stage model (ANN-based) for recording-level classification.
This ANN-based model consists of a single fully connected layer with a length equal to the input length, followed by a softmax layer. We have also experimented with using multiple layers of fully connected layers, adding activation functions between the layers, and obtained similar results. Therefore, we will not go into further detail on this.

Based on previous experimental results, our feature-accepting model consists of only one classification layer and a softmax layer. The classification layer is composed of a linear layer, with an input length corresponding to the flattened features and an output equating to the number of classification categories.

We conducted additional experiments to determine if a similar conclusion would be reached when the input consists of these extracted features. The results showed that introducing additional hidden layers and activation functions did not significantly impact the performance of the model. Therefore, we decided not to elaborate further on this aspect, as the additional complexity did not yield significant benefits in terms of model performance. This underscores the importance of balancing model complexity and performance in machine learning applications.

\begin{figure}[htbp]
\floatconts
  {fig:model diagram}
      {\caption{A typical deep learning first-stage model architecture: Firstly, the input passes through the model's feature extraction layer to be transformed into features. Then, these features pass through the classification layer to become logits. Finally, the logits are processed by the softmax layer to yield probability estimates.
		}}
  {\includegraphics[width=0.6\linewidth]{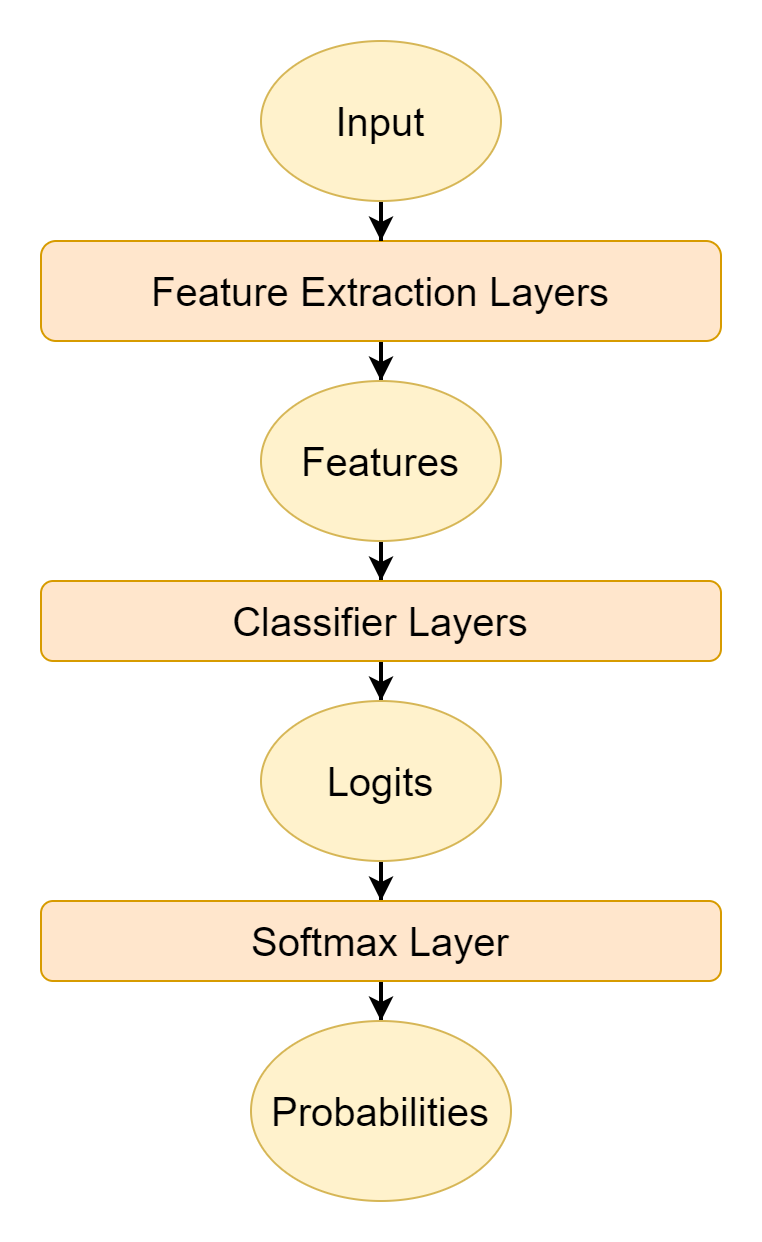}}
\end{figure}

\section{Results}
\subsection{Performance of Our Two-Stage Architecture}

On the TUAB dataset, we evaluated the performance improvement achieved by introducing our second-stage model, based on Artificial Neural Network (ANN) and XGBoost, compared with the first-stage model using 60 s windows (as in the original Deep4) and without overlapping. As shown in \figureref{fig:the second stage}, all our proposed machine learning arbitration methods outperform the baseline methods (`No arbitration', `Mean', and `Geomean'), thus indicating the second-stage model has a significant impact on the performance of the first-stage model. In particular, when using TCN and the second-stage model based on Raw (or Hybrid) and XGBoost, the average accuracy rate reached 99.0\% across 25 experiments. For the combinations of PaSST, Deep4, and ViT with XGBoost, the best performances were 98.7\%, 97.7\%, and 98.1\% accuracy, respectively, all substantial improvements upon the state-of-the-art in this task, as indicated in \tableref{tab:models}.

From Table \tableref{tab:models}, it is evident that, in comparison to previous models, we have addressed the pervasive issue of low sensitivity. We've enhanced the sensitivity to 98.1\%, which is also one of the metrics highly valued in clinical settings. This suggests that the majority of abnormal samples can now be accurately detected, bringing the model's performance closer to clinical expectations. Moreover, while past models have typically exhibited high specificity, our approach has achieved a perfect score of 100\%. This implies that samples without abnormalities are not misclassified.

Among the first-stage models considered, our implementation of ViT performed much worse than the others in most cases.
However, it is notable that the best-performing meta-model for this dataset, XGBoost with raw input, achieved similar accuracy regardless of the first-stage model choice.

\begin{figure*}[htbp]
\floatconts
  {fig:the second stage}
  {\caption{Performance comparison of single-stage and various two-stage architectures, all with a window length of 60~s, using the TUAB dataset. Each column represents a different arbitration method. Each marker type represents a different first-stage architecture. Each data point is the average accuracy across twenty-five experiments.
            Note that the accuracy of the `no arbitration' approach is calculated across all windows ($N = \num{57482}$), whereas the accuracy of the arbitration models is calculated across all recordings ($N=\num{2993}$)
  }}
  
  {\includegraphics[width=1\linewidth]{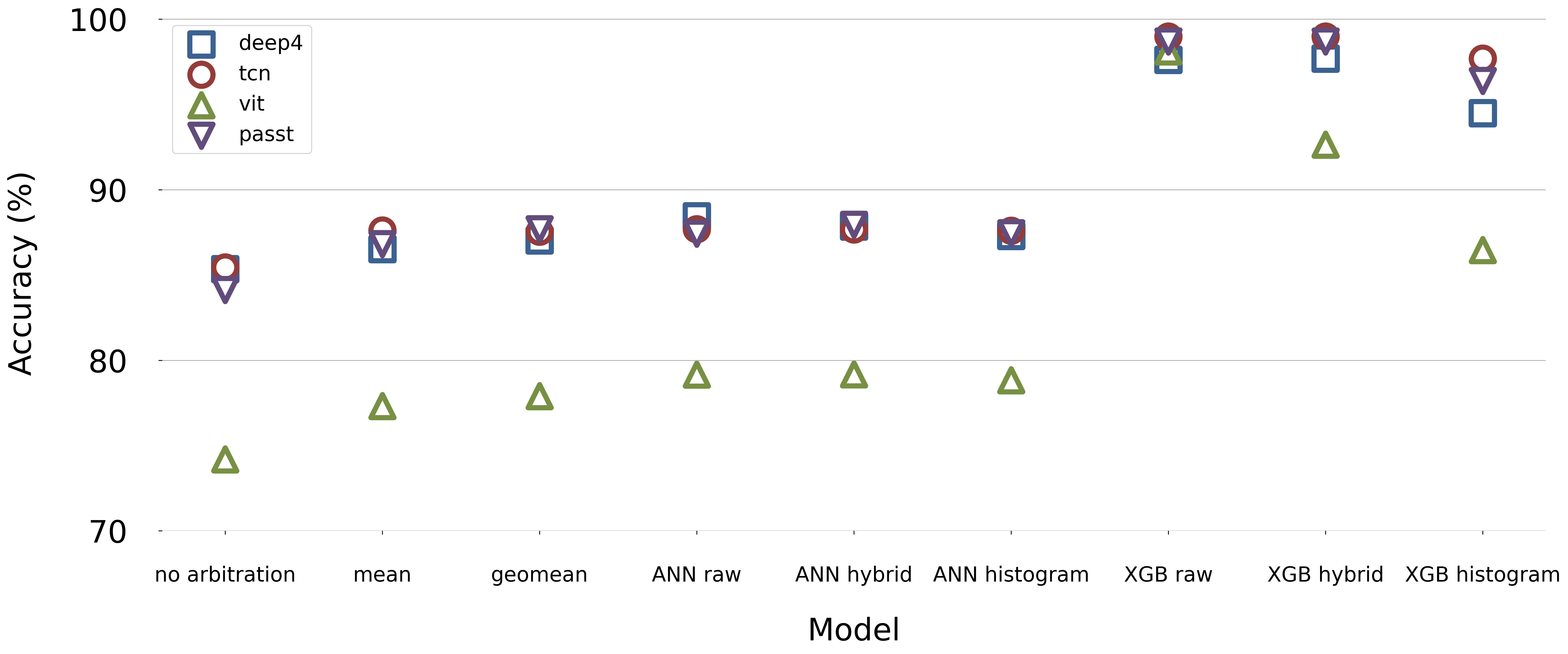}}
\end{figure*}

On the AutoTUAB dataset, as shown in \figureref{fig:AutoTUAB}, the ANN meta-model significantly improved the average accuracy of the model from 84.0 percent with `mean' arbitration to 86.9 percent, 
approaching the upper limit of the performance ceiling expected based on previously reported levels of inter-rater agreement. 
These results demonstrate the effectiveness of window stacking meta-models in improving the accuracy of abnormal EEG classification.

\begin{figure}[htbp]
\floatconts
  {fig:AutoTUAB}
  {\caption{Performance of multi-stage methods on AutoTUAB using Deep4 as the first-stage architecture with a window length of 60 s and no overlap. Each marker represents a single experiment and the dashed lines represent the mean accuracy of these 5 experiments. The third-stage model applied `mean' arbitration to the per-recording outputs of an ANN-based second-stage model, which took `raw' per-window probabilties as inputs.  }}
  {\includegraphics[width=1\linewidth]{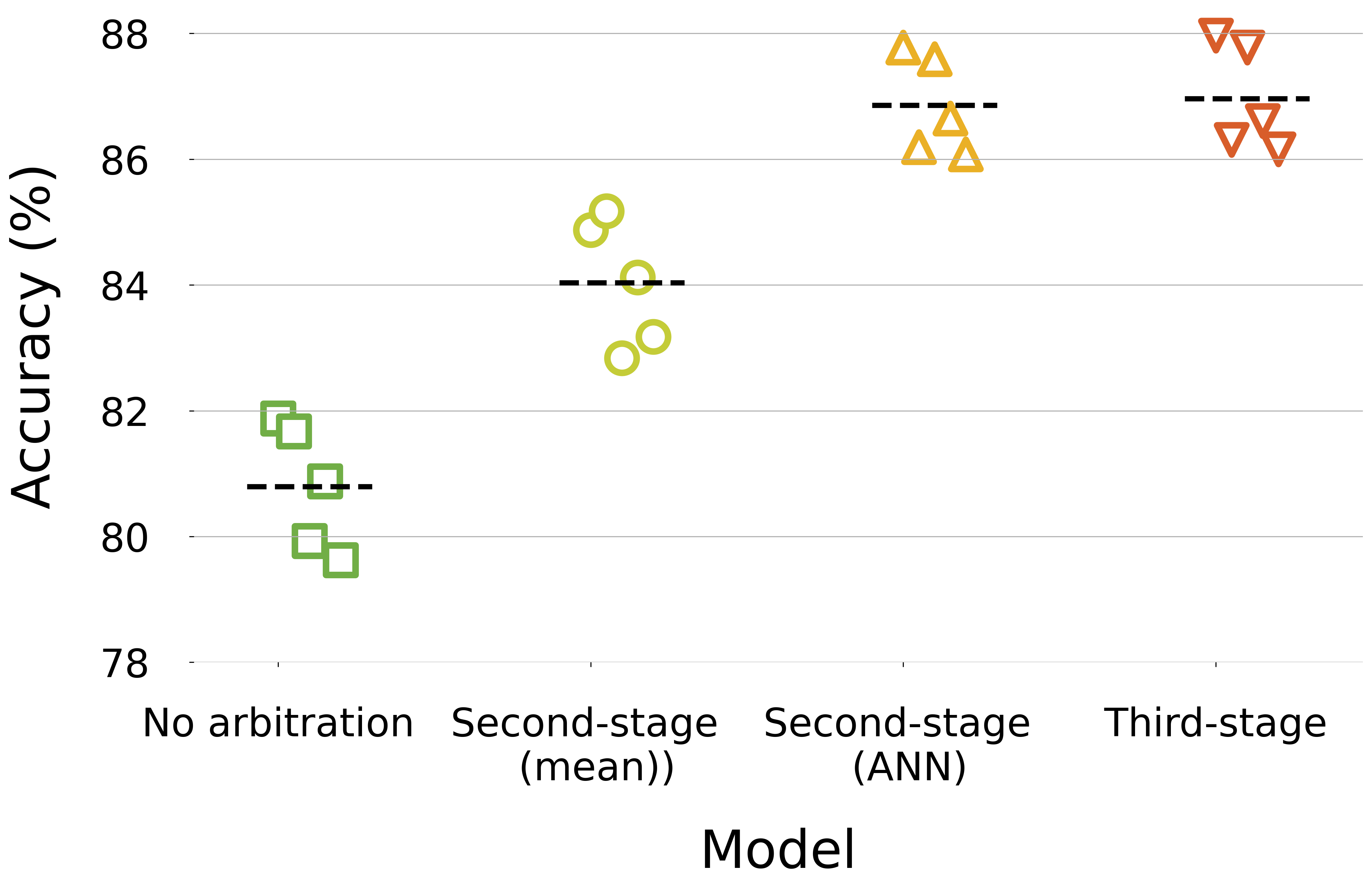}}
\end{figure}

\subsection{Effect of Window Length}

We conducted experiments to investigate the impact of window length on the performance of the various architectures. Our results, depicted in \figureref{fig:window length}, show that both the baseline model and the ANN-based model exhibit improved performance as window length increases. In contrast, the XGBoost-based model performs optimally with 180-second windows, after which its performance declines with longer windows.

\begin{figure}[htbp]
\floatconts
  {fig:window length}
  {\caption{Effect of window length on accuracy.
     In all cases, Deep4 is used as the first-stage model.}}
  {\includegraphics[width=1\linewidth]{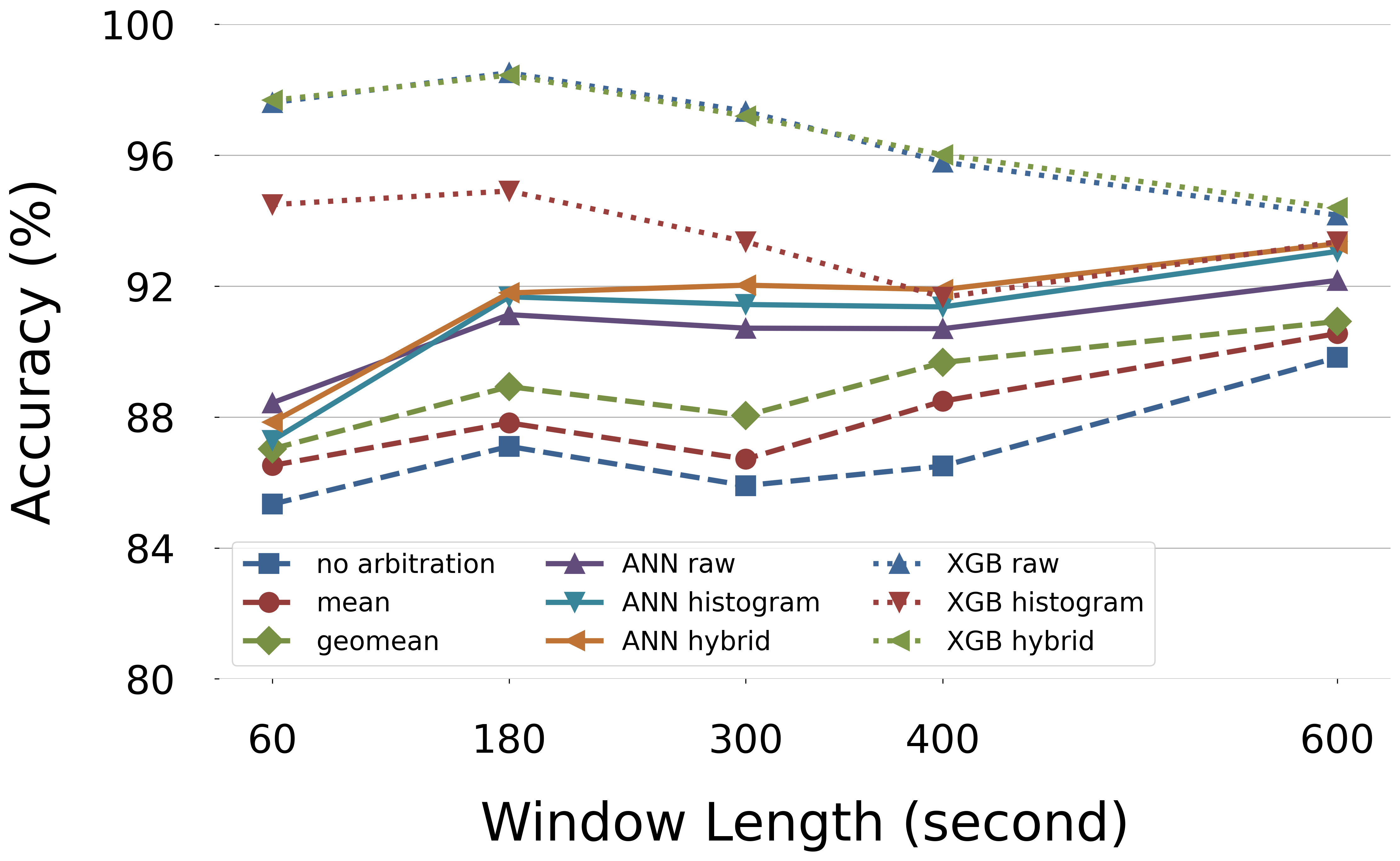}}
\end{figure}

\subsection{Overlapping}

As shown in the \figureref{fig:window stride}, the one- and two-stage architectures all benefit from overlapping, with greater overlap (smaller stride) tending to yield better accuracy. 
This trend is less pronounced in the case of the `raw' and `hybrid' XGBoost meta-models, perhaps due to a ceiling effect.

\begin{figure}[htbp]
\floatconts
  {fig:window stride}
{\caption{Effect of overlapping on accuracy (first stage = Deep4; window length = 60 s, dataset = TUAB).}}
  {\includegraphics[width=1\linewidth]{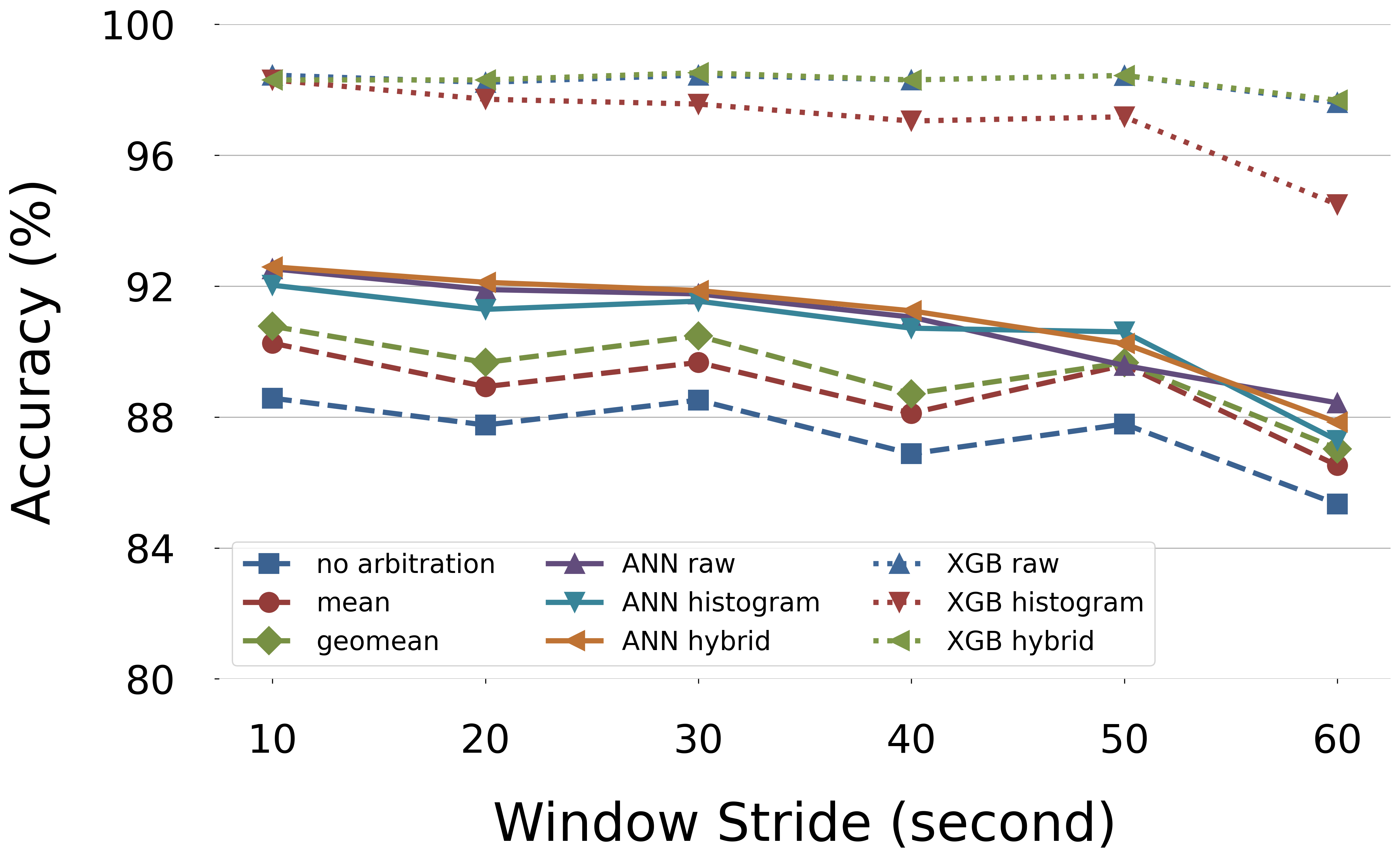}}
\end{figure}

\subsection{Different types of feature for the second-stage model}

As shown in the \figureref{fig:feature}, we tested the performance of the ANN and XGBoost meta-model when using different first-stage features. We found that when more detailed upstream features are used, the performance of the ANN-based model significantly improves relative to using the output of the classification layer. Features extracted from the convolutional blocks, the model's performance reached an average of 97.0\% over five experiments. 
On the other hand, using features still proves beneficial for the XGBoost-based arbitration model, albeit with a relatively small improvement. This is likely because XGBoost performs well with logits by default. When using features, the average accuracy of the XGBoost model reached 98.0\% over five experiments.

\begin{figure}[htbp]
\floatconts
  {fig:feature}
  {\caption{Comparison among different types of feature for the second-stage model (first~stage~=~Deep4; window~length~=~60~s, dataset~=~TUAB).
  }}
  {\includegraphics[width=1\linewidth]{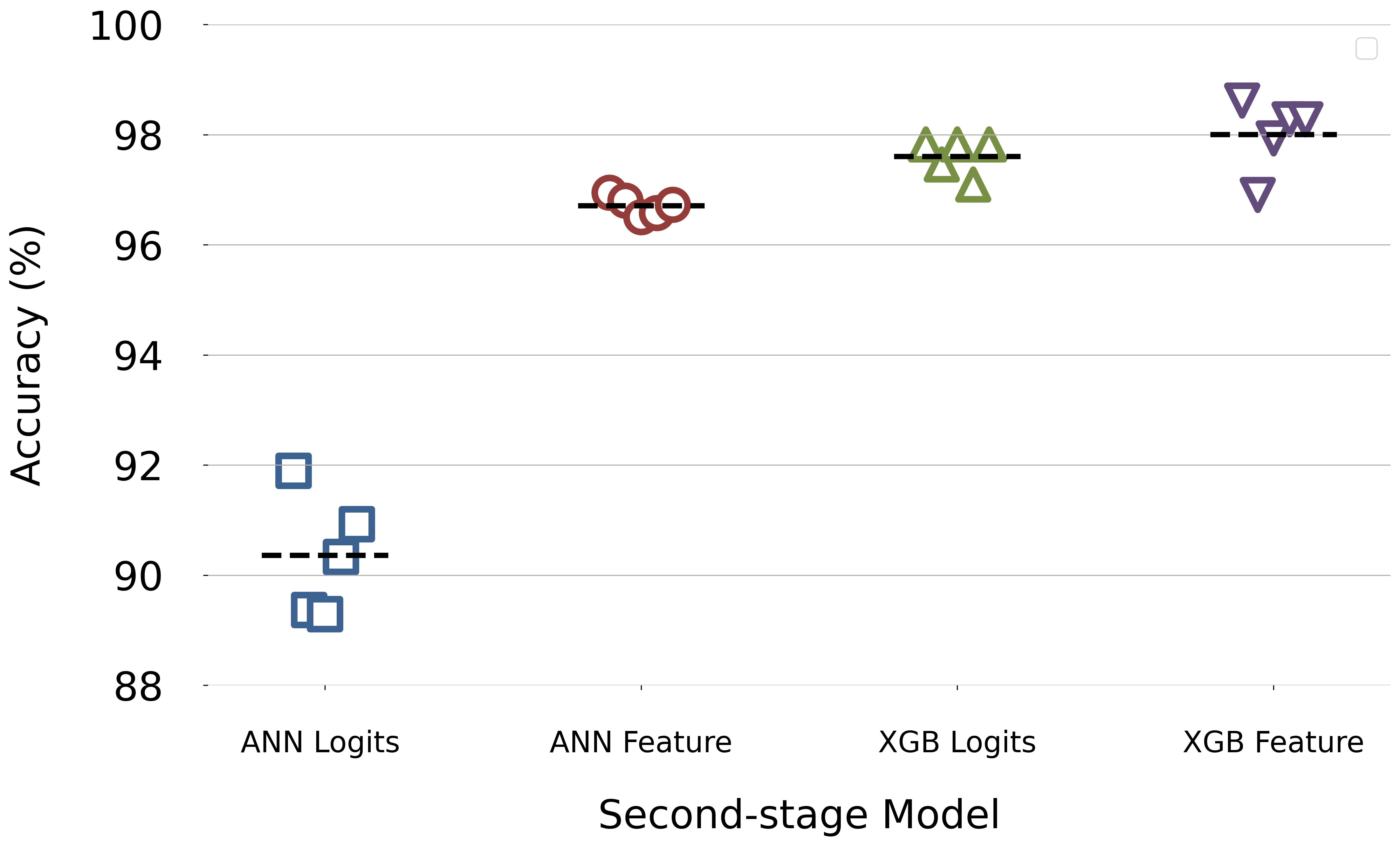}}
\end{figure}

\subsection{The third-stage model}
As illustrated in the \figureref{fig:3stage}, our session-level arbitration can further enhance the performance of the classification algorithms for sessions with multiple recordings, regardless of the second-stage architecture. The `geomean' method generally outperformed `mean', providing an improvement over `no session-level arbitration' in the range $1\%-2\%$.
After employing the third-stage models, our best average accuracy on the AutoTUAB dataset reached approximately 87.4 \%. 
Considering that this accuracy approximates the upper estimates of the performance ceiling imposed by interrater agreement, it is conceivable that greater improvement might be achieved by this method if the limits of interrater agreement can be overcome in the curation of future datasets. 

\begin{figure}[htbp]
\floatconts
  {fig:3stage}
  {\caption{The performance of the third stage model for the subset of AutoTUAB with multiple recordings per session (3725 out of 12240 sessions). Each marker type represents a different third-stage arbitration method. Each data point is the average of twenty-five experiments. In all cases, session-level arbitration methods (`mean' and `geomean') outperform `no arbitration'.).
  }}
  {\includegraphics[width=1\linewidth]{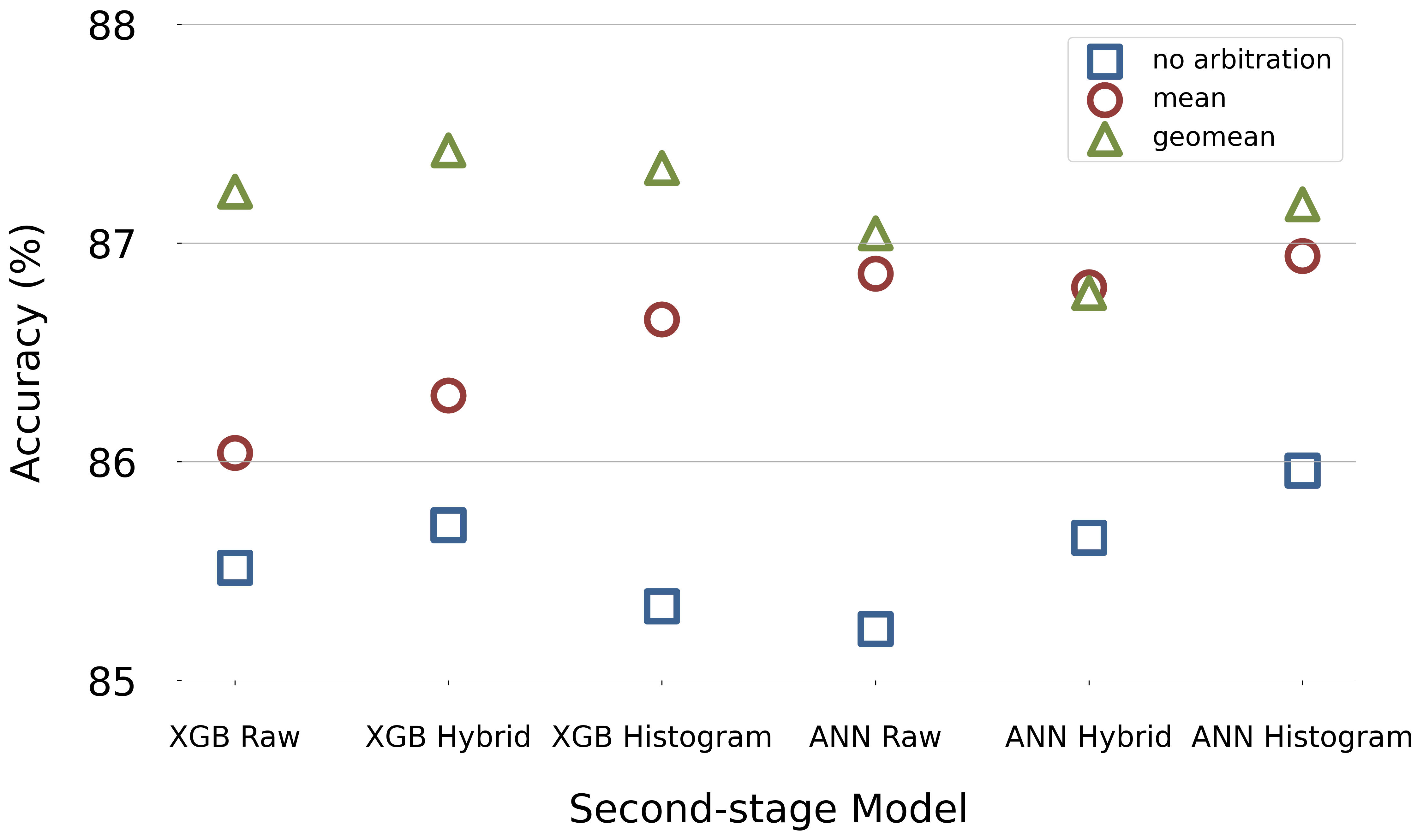}}
\end{figure}

\section{Discussion}

\subsection{Multi-Stage Model}

As demonstrated in \figureref{fig:the second stage,fig:AutoTUAB}, window stacking meta-models significantly enhance performance in clinical EEG classification.
Specifically, when using a 60-second window length, the XGBoost-based meta-model improved TCN's accuracy on the TUAB dataset from 85.3\% to 99.0\%.
Additionally, the effectiveness of our model was evident across multiple datasets, first-stage models, window lengths, and amounts of overlap. On the AutoTUAB dataset, our two-stage model improved Deep4's accuracy from 80.8\% to 86.9\%, while the three-stage model further improved the accuracy to 87.0\%. 

We propose that our multi-stage models are effective for two reasons. 
\begin{enumerate}
\item 
They extend the scope of the model to consider a full recording/session without substantially increasing the number of parameters in the model. 
This avoids an increase in computational requirements and the amount of data required to train the model effectively, which is particularly important while high-quality, widely available, clinical EEG datasets are relatively scarce. 
The deep first-stage models can be trained across a large number of windows, whereas the relatively simple meta-models can be adequately trained on the smaller number of available recordings. 
For end-to-end training of a single model with equivalent scope, the number of samples available for training all parameters would be limited to the number of recordings.  

\item 
The meta-models directly address the issue of misleading redistribution of labels (e.g. when an apparently normal window inherits a label of `abnormal' from its parent recording) by optimising the arbitration stage, which has been included but received very little attention in past studies.
Our results show that simple arbitration (e.g. `mean') is better than no arbitration - it gives the system robustness to occasional misleading redistribution of labels and allows confidence to be aggregated across repeated measures - but a machine learning meta-model can improve substantially over simpler non-parametric approaches.
\end{enumerate}

For the second-stage model, XGBoost outperforms ANN-based models and non-parametric methods, suggesting that a strong meta-model must learn a complex mapping relationship. This may be due to XGBoost's well-established superiority over ANNs for tabular data. 

Session-level (third-stage) arbitration was found to provide additional benefit for cases with multiple separate recordings, although such cases represent only a small portion of the AutoTUAB dataset (1.07 recordings per session, on average).
Further benefits of this approach may be realised in other datasets, such as the recently released Harvard Encephalography Database (approximately 200,000 files from 63,000 patients) \citep{zafar_harvard_2023}.

Finally, we also demonstrated that using features from the intermediate layers of the first-stage model could enhance the performance of the meta-model.
We have not yet explored the application of an XGBoost meta-model to the intermediate features of our best performing first-stage architecture, TCN, due to computational expense. 
Based on \figureref{fig:feature} we expect that it may yield further improvement on our state-of-the-art accuracy for TUAB (TCN-XGBoost-raw), but the improvement would be marginal due to ceiling effects.
We intend to focus first on the exploration of broader datasets before evaluating this combination.

\subsection{Overlapping and Window Length}

The results in the \figureref{fig:window length} indicate that the window length has varying impacts on the performance of the baseline model, the ANN-based model, and the XGBoost-based model. For the baseline model and the ANN-based second-stage model, longer windows result in better performance, whereas XGBoost performs better with smaller windows. We posit that longer windows can effectively alleviate the issue of misleading redistribution of labels by reducing their prevalence, since the longer windows have greater scope to include phenomena relevant to determining the label. 
However, the powerful fitting and mapping capabilities of XGBoost for tabular data allow it to effectively address the problem of misleading redistribution of labels through arbitration, which is further improved by using smaller windows to increase temporal resolution. 
The optimal window size is a trade-off between temporal resolution and scope, given that XGBoost's peak performance occurs with a window size of 180 s.

Furthermore, \figureref{fig:window stride} demonstrates that overlapping can improve the performance of both XGBoost and ANN when using a 60 s window. 
This may be because increasing the sample size of the first-stage model leads to more stable training and improves the utilization of the second-stage model's temporal resolution. 
However, the effect on XGBoost is less pronounced, potentially because its performance is already close to the upper limit.
The overlap experiments focussed on Deep4 as the first-stage model to limit computational expense, and the best variant using Deep4 with overlap still did not perform as well as TCN with no overlap.

\subsection{Performance Ceilings of Machine Learning Models for Clinical EEG Abnormality Classification}
\label{sec:ceilings}

In our previous research, we discussed how model performance might be influenced by inter-rater agreement \cite{zhu2023scope}. Our results on TUAB (e.g. \figureref{fig:the second stage}) demonstrate for the first time that this ceiling does not apply to TUAB. 
This can be explained based on how samples were selected and labelled for TUAB.
As noted in the dataset's `readme' document, samples were selected for inclusion ``based on their relevance to the
normal/abnormal detection problem - whether they display some
challenging or interesting behavior.''
In the more recently released v3.0.0 of TUAB, the curators report that their labels on the TUAB test set achieve 100 percent agreement with the clinicians' reports.
Furthermore, the labelling was performed by consensus among a panel of interpreters \cite{lopez2017automated}.
Although different raters may give different labels to the same EEG signal, a machine learning model can learn to replicate the judgement of one rater (or team of raters, as used in TUAB).
Normal levels of inter-rater agreement clearly do not apply to TUAB.

In contrast, the selection criteria for AutoTUAB are more inclusive;
the minimum recording length is lower and, more importantly, there is no direct selection based on the features observable in the recordings.
The natural language processing used to extract labels from the clinical reports may impose some selectivity. 
Only the labels attributed with very high confidence were used, resulting in the exclusion of 13 percent of the database.
Also, recordings without the full 10-20 electrode montage were excluded, like in TUAB, which may bias the dataset against some phenotypes.
Nonetheless, the influences of these selectors on inter-rater agreement within the dataset are expected to be relatively small.
We expect the performance ceiling on AutoTUAB to be similar to inter-rater agreement levels in the literature, although it may be slightly higher. 
As shown in \figureref{fig:AutoTUAB}, our models' performance on AutoTUAB approaches the upper end of previously reported inter-rater agreement levels (82-88 percent).

Now that machine learning approaches can, in some sense, match human expert performance in this task, future work should include the curation of datasets that combine a diverse range of human expert judgements on individual samples and/or data on alternative clinical outcomes/metrics to corroborate and optimise label accuracy.

\subsection{Algorithmic Explainability}

\subsubsection{Examining Error Cases}

Our best performing two-stage architecture produced only two or three errors in each of five experiments on the TUAB test set. 
All of these were false negatives from the same three recordings. 
To better understand our model's behaviour, these three cases were manually reviewed by a clinical neurophysiologist. 
This analysis revealed that the evidence of interictal epileptiform discharges (IEDs) was equivocal in all three cases, although the `abnormal' label was further supported by evidence of slowing in two cases and photoparoxysmal responses in a third. 
It is conceivable that the first-stage model placed high importance on IEDs, leaving it prone to error in cases where these features are debatable. 
This brief analysis does not offer a conclusive explanation of these error cases, but it increases our confidence that the errors are due to plausibly challenging cases and not due to extreme outliers or anomalies in the dataset.

\subsubsection{Window Importance Distribution}

In the interest of algorithmic explainability, we examine the importance attributed by the Deep4-ANN-Raw meta-model to each window position based on the model's weights, as described in Equation \eqref{eq:importance}. 

\begin{figure}[htbp]
\floatconts
  {fig:data and weight}
  {\caption{Examining the influence of window position on model behaviour in terms of a) weighting of each window in the `Deep4-ANN-raw' model and b) per-window accuracy of the first-stage model  c) recording length histogram (60 s window length, no overlap).
}}

    \subfigure[Weight for each window in second-stage model on AutoTUAB]{\label{fig:AutoTUAB_weight_60s}%
		\includegraphics[width=1\linewidth]{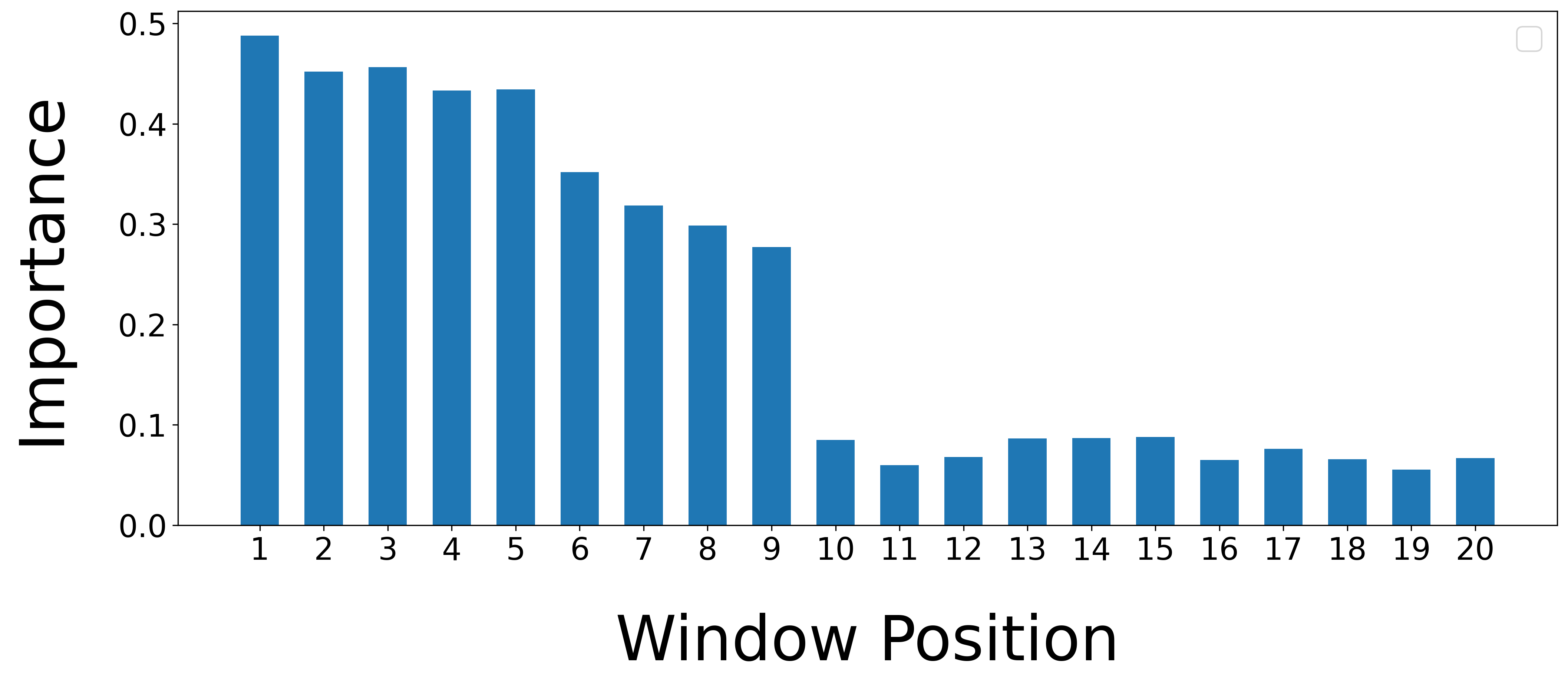}}
	
    \subfigure[Accuracy of first-stage outputs on AutoTUAB]{\label{fig:AutoTUAB_data_60s}%
      \includegraphics[width=1\linewidth]{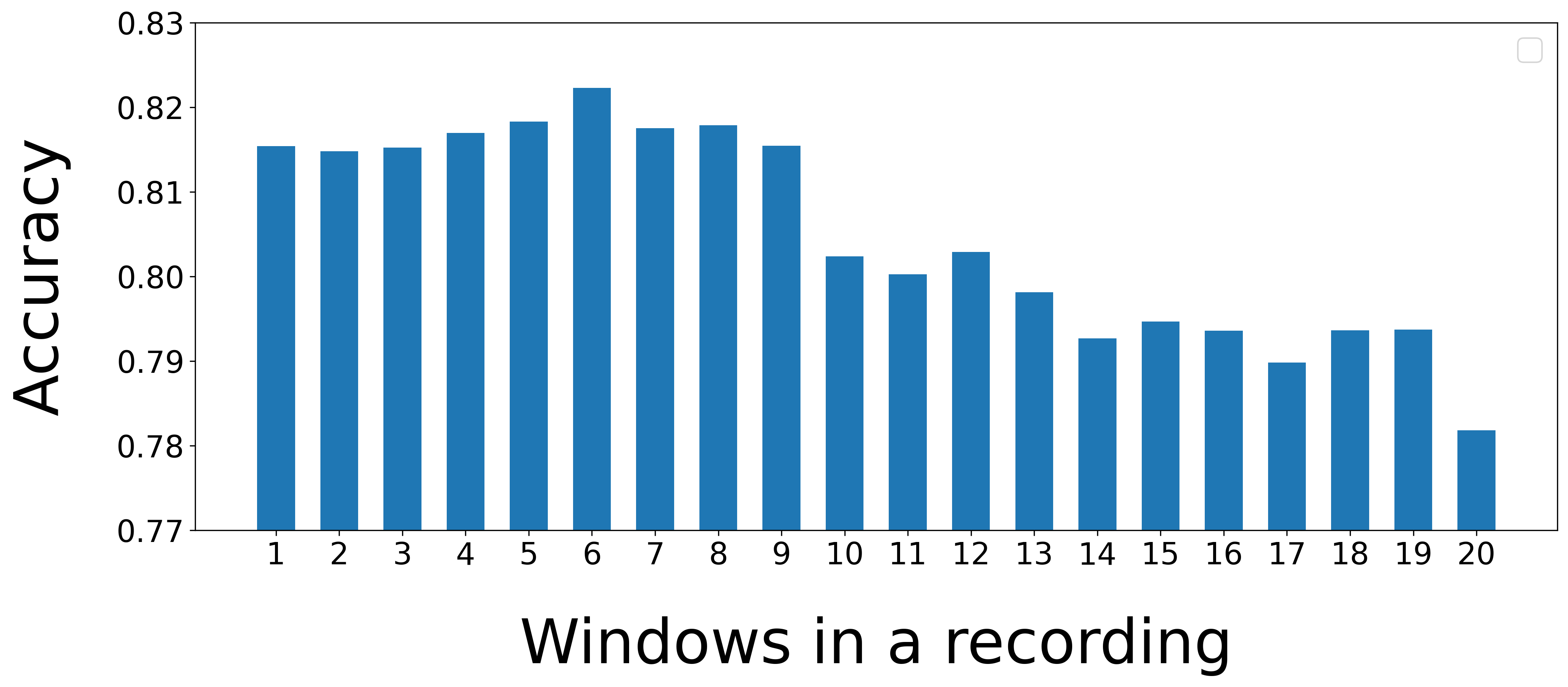}}%
      
	\subfigure[Recording length histogram of AutoTUAB dataset]{\label{fig:AutoTUAB_valid_lens_60s}%
      \includegraphics[width=1\linewidth]{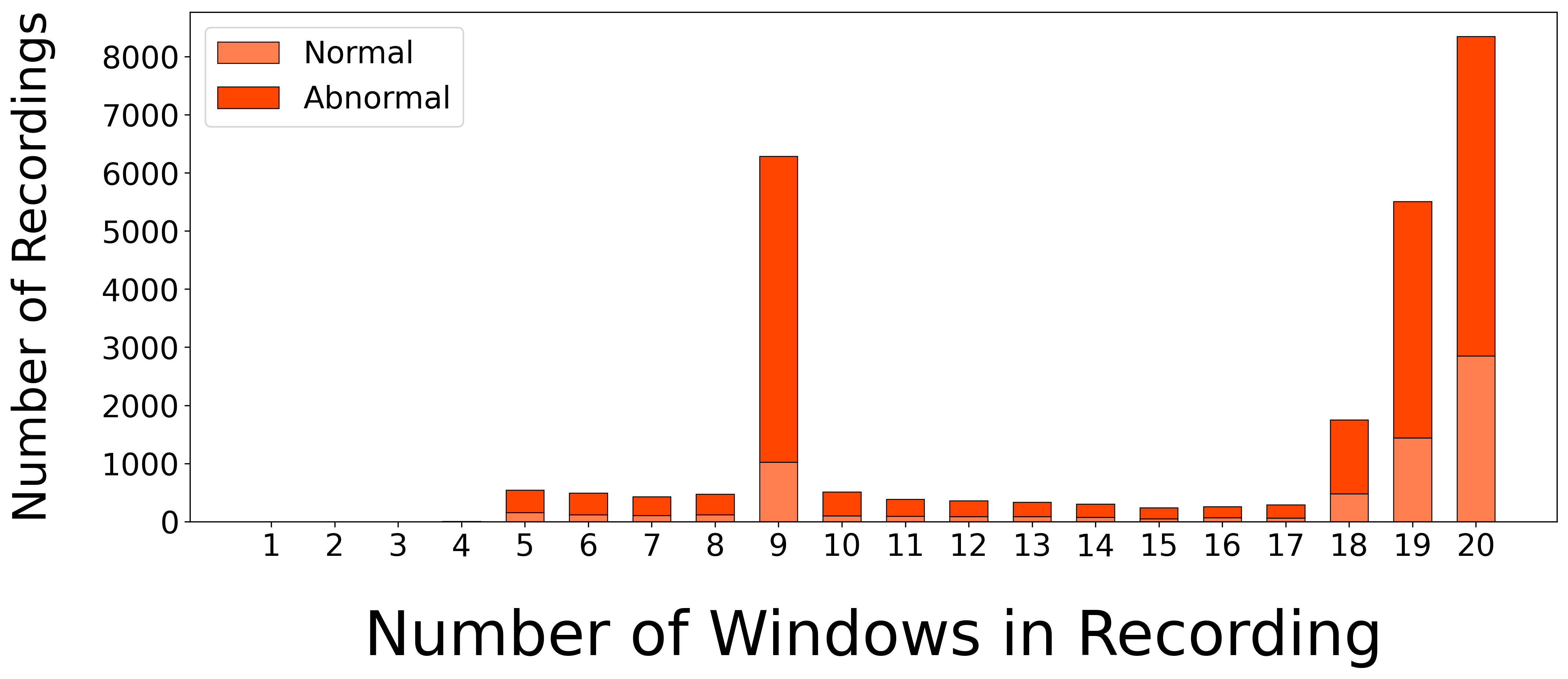}}

\end{figure}

In \figureref{fig:AutoTUAB_weight_60s} the importance exhibits a downward trend across the first nine windows (the first ten minutes of the recording, with the first minute is routinely excluded).
This pattern is broadly consistent with previous observations that strong classifier accuracy can be obtained using relatively short recording lengths \citep{schirrmeister2017deep}.

After the ninth window, this trend is interrupted by a sharp decrease, and the importance of subsequent windows is consistently relatively low.
\figureref{fig:AutoTUAB_data_60s} shows that the accuracy of the first-stage model decreases after the ninth window and exhibits a steady downward trend thereafter.
This phenomenon can be explained by the distribution of recording lengths depicted in \figureref{fig:AutoTUAB_valid_lens_60s}.
A notably large portion of recordings in AutoTUAB have an original length of ten minutes, perhaps reflecting a recording protocol specific to a particular clinical setting.
This can be presumed to be an artefact of variations in the data pruning practice adopted when archiving the clinical recordings \cite{picone_tueg_2023}.
The first-stage model therefore receives fewer training examples from later windows. 
The fact that this results in a small drop in accuracy suggests that there are subtle differences between these and earlier windows.
The `raw' meta-model inputs for the shorter recordings are padded with zeros (implying no evidence of abnormality) in place of the `missing' windows.
The sharp drop in importance attributed to later windows indicates that the zero-padding makes these window positions less informative to the ANN than those that are consistently occupied by genuine data. 

Given that the ratio of abnormal to normal recordings appears to be higher in the shorter recordings (\figureref{fig:AutoTUAB_valid_lens_60s}) in AutoTUAB, it is conceivable that the meta-model would learn to `cheat' by adopting a bias towards `abnormal' whenever it detects zero-padding in the later windows.
However, any such bias would not improve performance in the TUAB test set in which all recordings have a length greater than 15 minutes, so it cannot explain the key results of this paper.
Nonetheless, future exploration of window-stacking meta-models should carefully consider such characteristics of the data used and the potential influence on model behaviour.

As illustrated in  \figureref{fig:xgboost shap}, we employed SHAP values to delve into the influence of different windows on the decision-making of an XGBoost meta-model.
Overall, it is evident that earlier windows tend to be more important in determining the meta-model's predictions. 

Closer inspection yields further insight into some of the strategy adopted by the meta-model.
For example, the distribution of SHAP values is asymmetrical; dense red clusters on the left of \figureref{fig:xgboost shap} indicate that a window perceived to be highly abnormal by the first-stage model can strongly influence the meta-model towards a decision of `abnormal', whereas the influence of an individual `normal' seeming window is more variable (less densely clustered).
This observation suggests that the meta-model has adopted the intuitive strategy alluded to in \sectionref{sec:intro}, which `mean' arbitration neglects: a clear-yet-isolated abnormality can be taken as grounds for classifying the whole EEG as `abnormal', even though the opposite conclusion cannot be drawn from a lone normal window.

In summary, the window stacking meta-models tend to attribute greater weighting to earlier portions of the recording, corroborating previous indications that earlier portions are particularly valuable to the classification \citep{schirrmeister2017deep}.
The meta-models adopt strategies that are consistent with human intuition, while also being optimised for the available training data.
However, they also learn strategies particular to subtle aspects of the distribution of data in the training set.
This may compromise the generalisability of a trained meta-model to other datasets.
Such considerations should be factored into the further development of window stacking meta-models and of new datasets for machine learning EEG classification.

\begin{figure}[htbp]
\floatconts
  {fig:xgboost shap}
  {\caption{SHAP Summary Plot for Model Predictions. This visual representation showcases the importance and impact of different windows on an XGBoost meta-model's predictions, based on the window's position within a recording. 
  Window positions are listed vertically, ranked from top to bottom based on their importance. 
  Each dot represents a SHAP value for a specific instance of a window position (i.e. the influence of the first-stage model output for that window in a single recording). 
  The colour indicates the first-stage model's output value (i.e. the predicted probability of abnormality).
  }}

  {\includegraphics[width=1\linewidth]{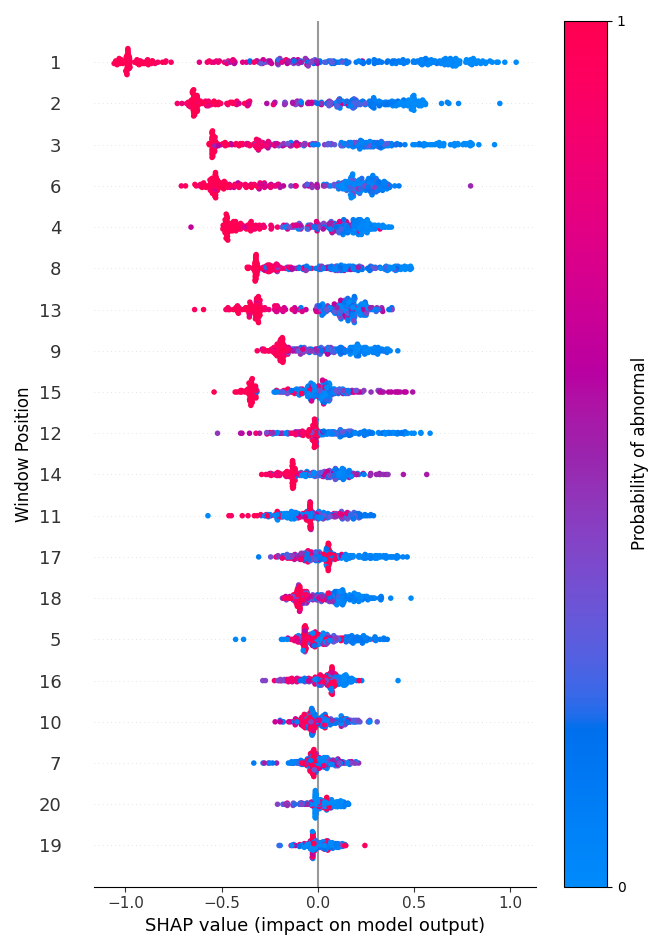}}
\end{figure}

\subsection{Future Work}

As discussed in \sectionref{sec:ceilings}, our window stacking meta-models demonstrate classification accuracy at or very close to the performance ceilings for these datasets, but further improvements in real-world performance may be enabled by the curation and sharing of new datasets with more reliable ground truths.
Once such datasets are widely available, further architectural improvements may be sought in the first stage or in second- or third-stage models (meta-models).
Investigating the effect of session-level arbitration (our third-stage model) on a dataset with more recordings in each session may reveal greater benefits from this approach.

The window stacking meta-model concept can also be tested on different task types, such as multi-classification tasks or other time-series classification problems, for insight into its broader applicability.

Finally, the multi-stage model has limitations in terms of gradient transfer between the models. As more clinical EEG data with suitable labels is made available, researchers will likely take advantage of this through further end-to-end training of larger single-stage models. 
Additionally cross-domain pre-training, self-supervised learning of EEG data, or multi-modal learning of EEG data could be explored, to gain further improvements in first-stage performance with available data. 
These first-stage improvements would remain compatible with a multi-stage approach.
Overall, future research in this area has the potential to improve the accuracy and scalability of EEG classification models.

\section{Conclusion}

With the introduction of window-stacking meta-models, we have substantially improved upon previous state-of-the-art performance in clinical EEG classification as measured against the TUAB dataset. 
This approach optimises an important aspect of conventional machine learning EEG classification pipelines that has been given little previous attention.
By improving the overall model's robustness to the imperfect nature of window labels inherited from their parent recording, window-stacking meta-models improve the accuracy --- and particularly the sensitivity --- of recording- or session-level classification.
Furthermore, they do so without significantly increasing computational expense or the required volume of training data.
Machine learning classifiers can now match human performance in the accuracy of abnormal EEG classification.
Future work should give particular focus to translating this work into clinical practice, as well as curating and sharing new datasets (or augmenting existing ones) to enable further improvement beyond typical human performance.

\acks{This work was supported by a PhD studentship funded by Southmead Hospital Charity and the University of the West of England.}

\bibliography{jmlr-sample}

\section*{Declaration of Interest}
Declarations of interest: none.

\end{document}